\documentclass[twocolumn, prl, aps]{revtex4-2}
\usepackage[english]{babel}
\usepackage[utf8]{inputenc}
\usepackage{lmodern}

\usepackage{dcolumn}
\usepackage{amssymb, amsmath}
\usepackage{graphicx}
\usepackage{graphics}
\graphicspath{{figures/}} 
\usepackage{bm}
\usepackage{xcolor}
\usepackage{nicefrac}
\usepackage{hyperref}
\usepackage{booktabs}
\usepackage{multirow}
\usepackage{microtype}
\usepackage{braket}
\usepackage{physics}

\begin{document}
\title{Efficient spin accumulation carried by slow relaxons in chiral tellurium}
\author{Evgenii Barts}
\author{Karma Tenzin}
\author{Jagoda S\l awi\'{n}ska}
\affiliation{Zernike Institute for Advanced Materials, University of Groningen, Nijenborgh 3, 9747 AG Groningen, The Netherlands}
\date{\today}

\begin{abstract}
Efficient conversion between charge currents and spin signals is crucial for realizing magnet-free spintronic devices. However, the strong spin-orbit coupling that enables such a conversion, causes rapid relaxation of spins, making them difficult to transport over large length scales. Here, we show that spin-momentum entanglement at the Fermi surface of chiral tellurium crystals leads to the appearance of slow collective relaxation modes. These modes, called relaxons, resemble the persistent spin helix, a collective spin-wave excitation with extended lifetime observed in quantum wells. The slow relaxons dominate the electrically generated spin and orbital angular momentum accumulation in tellurium and make it possible to combine a very high 50\% conversion efficiency with a long spin lifetime. These results, obtained from the exact solution of the Boltzmann transport equation, show that chiral crystals can be used for highly efficient generation and transmission of spin signals over long distances in spintronic devices. 
\end{abstract}

\maketitle
\section*{Introduction}

Spintronic devices that make use of both the charge and spin of electrons can provide high-performance and energy-efficient solutions for electronics~\cite{flatte}. However, progress in the field is hindered by the insufficient compatibility between semiconductors and ferromagnets used for spin injection and detection~\cite{bart_obstacle}. An alternative approach without ferromagnetic electrodes relies on the generation of spin signals via charge currents using charge-to-spin conversion (CSC) mechanisms~\cite{she, ganichev_nature}. Nevertheless, the conversion efficiency in these processes typically does not exceed a few percent, which is far below the values required for practical devices and circuits~\cite{intel}. Moreover, the strong spin-orbit coupling (SOC) that makes CSC phenomena detectable has a detrimental effect on the generated spin signals. It creates an effective momentum-dependent magnetic field that acts on electrons traveling through the crystal, causing precession of the spin and its dephasing upon scattering. Thus, using charge currents for efficient generation of spin signals that could survive over large distances remains a fundamental trade-off. 

\begin{figure*}
	\centering
	\includegraphics[width=\linewidth]{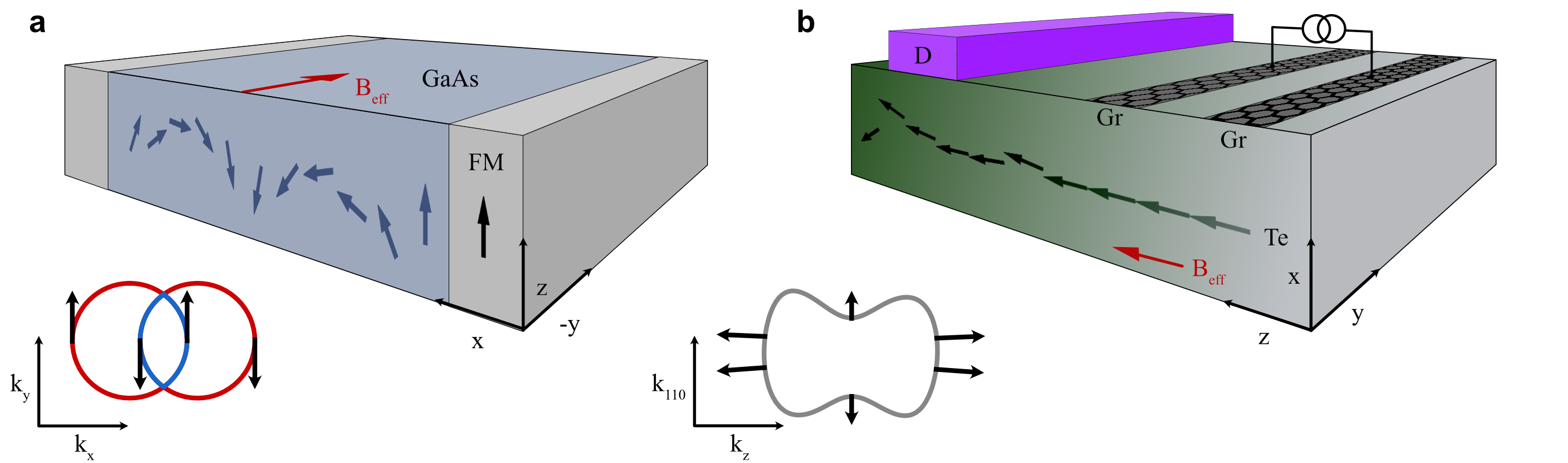}
	\caption{\textbf{$\mid$ Spin transport in GaAs quantum wells and in chiral Te crystals. a} Scheme of the coherent spin precession in GaAs with equal Rashba and Dresslehaus parameters ($\alpha_R=\alpha_D$). The inset illustrates the spin polarization of bands in this model resulting in a persistent spin texture along $k_y$.  The spins are injected from the ferromagnet and travel along the $x$ direction over a large distance. The effective spin-orbit magnetic field ($B_{eff}$) along the $y$ axis causes coherent spin precession in the $xz$ plane, independently of scattering events. \textbf{b} Example of a device for generating, transmitting and detecting spin accumulation in Te. The electric current is applied along the $z$ axis and the induced spin accumulation along the current direction is transmitted toward the detector. Beyond the region with the electrodes, the spin accumulation slowly decays. The inset schematically illustrates the 
topmost valence band of Te enabling the efficient spin transport. 
    }
	\label{fig:PSHintro}
\end{figure*}

The idea of a peculiar spin precession mode called a persistent spin helix (PSH) was suggested as a remedy to protect spins from decoherence in the presence of strong SOC~\cite{sfetloss}. It relies on the engineering of spin-orbit interaction via external tuning of the system's symmetries. When inversion symmetry is lacking, SOC splits energy states, manifesting as a form of Zeeman interaction with a momentum-dependent magnetic field. Depending on symmetries, the spin-orbit field acting on electron spin can be of Rashba or Dresselhaus type:
\begin{equation}
        H_{\rm SOC} = \alpha_{\rm R} 
    \left(k_x \sigma_y - k_y \sigma_x
    \right)+ \alpha_{\rm D} 
    \left(k_x \sigma_x - k_y \sigma_y
    \right).
\end{equation}
If their strengths are equal ($\alpha_{\rm R} = \alpha_{\rm D}$), SOF becomes unidirectional and momentum-independent, yielding a configuration known as a persistent spin texture (PST)~\cite{berenevigpst}. In real space, the spin of a moving electron is subjected to a controlled precession around the unidirectional spin-orbit field, which results in a spatially-modulated periodic mode protected against decoherence.  Thus, PST allows for a spin-wave collective excitation characterized by an infinite lifetime, PSH, sketched in Fig.~\ref{fig:PSHintro}{\bf a}.  PSH was experimentally observed in GaAs quantum wells with balanced Rashba and Dresselhaus SOC~\cite{koralek}, and its real-space helical pattern was later imaged optically~\cite{Walser_2012}. Despite its fundamental importance, PSH was not commonly explored for devices due to the need for fine-tuning of SOC parameters and temperature limitations of quantum wells. 

Recently, the same mechanism for generating PST has been discovered in orthorhombic crystals, where the interplay of crystal symmetries naturally ensures equal strengths of Rashba and Dresselhaus interactions, enforcing a unidirectional spin polarization of states in momentum space~\cite{tao}. Several materials with PST were theoretically predicted~\cite{carmine, snte, rondinelli}, and suggested as ideal candidates for robust spin transport at room temperature. Surprisingly, PSH in solid-state materials has rarely been explored experimentally~\cite{perovskite_photonics, spin_dephasing}. This is due to the fundamental limitation that in systems with equal Rashba and Dresselhaus parameters, spin accumulation cannot be induced by an electric current~\cite{Duckheim2010}, so spins still need to be injected using ferromagnetic electrodes or excited optically. This raises questions about the compatibility of PSH with all-electrical generation, manipulation, and detection of spin signals. Can we find materials with strong SOC enabling both the efficient electrical generation of spin accumulation and long-range spin transport via a mechanism similar to PSH? 

In this Article, we show that bulk crystals of tellurium (Te) enable charge-to-spin conversion with unprecedented efficiency, reaching 50\%. In contrast to systems with equal Rashba and Dresselhaus parameters, spin generation in Te is possible via the collinear Rashba-Edelstein effect (REE), whereby an applied charge current induces accumulation of spins parallel to the current direction~\cite{arunesh_npj, nanowires}. When the current flows along Te chains, a persistent spin texture aligned in the same direction boosts the spin accumulation efficiency and partially suppresses the back-scattering of electrons, as presented in the inset of Fig.~\ref{fig:PSHintro}{\bf b}. This results in an extended spin lifetime, making Te suitable for both spin generation and transport in devices similar to the one shown in Fig.~\ref{fig:PSHintro}{\bf b}. To quantify these effects, we use the concept of relaxons allowing for exact solutions of the Boltzmann transport equations. Our calculations of current-induced spin and orbital angular momentum accumulation quantitatively agree with nuclear magnetic resonance (NMR) experimental data~\cite{tellurium_main}. Furthermore, we reveal that the relaxation of the non-equilibrium electron population is characterized by two collective relaxation modes, but only the slower mode, resembling PSH in quantum wells, carries spin momentum and governs the magnitude and lifetime of spin accumulation. Therefore, the generated spin signals relax at rates slower than the mean electron momentum relaxation rate, supporting long-range spin transport.

\section*{Results and Discussion}
\subsection*{Exact Boltzmann transport approach}

To quantify the current-induced spin accumulation in tellurium crystals, we use the semi-classical Boltzmann equation~\cite{abrikosov1988fundamentals}. It describes the time evolution of the electron distribution function, $f_{\bm k}$, under external stimuli:
\begin{equation}\label{eq:Boltzmann}
    \frac{\partial f_{\bm k}(t)}{\partial t} +  (-e) \bm E  \cdot \frac{\partial f_{\bm k}(t)}{\hbar\,\partial \bm k}
 = \left(\frac{\partial f_{\bm k}}{\partial t}\right)_{\rm col},
\end{equation}
where $\bm k$ is the reciprocal vector, $\bm E$ is the applied electric field, and $(-e)<0$ is the electron charge. The collision integral on the right-hand side of Eq.~\eqref{eq:Boltzmann} accounts for scattering processes with impurities. While the Boltzmann equation allows us to compute charge and spin transport properties, it is challenging to solve without approximations.

A common approach to solving this equation is the relaxation time approximation~\cite{abrikosov1988fundamentals}, given by 
$\left(\frac{\partial f_{\bm k }}{\partial t}\right)_{\rm col} =- {\delta f_{\bm k }}/{\tau}
$. Although this approximation often captures the key properties of transport phenomena, it has limitations~\cite{Vyborny2009}. 
Specifically, it predicts an exponential decay of the deviation from the equilibrium Fermi-Dirac distribution: $
\delta f_{\bm k}(t) = f_{\bm k}(t) - f_{\bm k}^{(0)} \sim 
e^{-t/\tau}$,
which can oversimplify the dynamics. Meanwhile, determining whether one or multiple characteristic time scales describe the dynamics of a system is more complex.
This issue becomes more pronounced in materials where the spin relaxation time significantly exceeds the mean electron scattering time.
Thus, the constant relaxation time approximation is particularly limited when describing materials with extended spin lifetimes.

To address this, we solve the Boltzmann equation exactly using a microscopic form of the collision integral~\cite{abrikosov1988fundamentals}:
\begin{equation}    
\left(\frac{\partial f_{\bm k}}{\partial t}\right)_{\rm col} = - \sum_{\bm k^{\prime} } W_{\bm k \bm k^{\prime} }\left(
f_{\bm k} -
f_{\bm k^{\prime}}\right),
\end{equation}
where the scattering probability, $W_{\bm k \bm k'}$, is given by Fermi's golden rule:\newline
$
W_{\bm k \bm k'} = \frac{2\pi}{\hbar} |\langle \bm k' | H_{\rm int} | \bm k \rangle|^2 \delta(\varepsilon_{\bm k'} - \varepsilon_{\bm k})
$. 
Here, $\varepsilon_{\bm k}$ is the band dispersion, and $H_{\rm int}(\bm r)=U_{\rm imp}\sum_a \delta(\bm r - \bm r_a)$ describes the interaction with static impurities, modeled by delta-function short-range potentials of strength $U_{\rm imp}$ and randomly distributed with a low density $n_{\rm imp}$. Further details of the model implementation can be found in Methods. Importantly, the scattering amplitudes $\langle \bm k' | H_{\rm int} | \bm k \rangle$ depend on the spin, orbital, and sublattice degrees of freedom encoded within the states $| \bm k \rangle$, which enables accurate calculations of transport coefficients even in systems with strong entanglement of these degrees of freedom with momentum $\bm k$.

By introducing an ansatz $\delta f_{\bm k}(t) =   e^{-t/\tau} \mathcal{V}_{\bm k}$, we transform the Boltzmann equation at $\bm E = 0$ into an eigenvalue problem:
\begin{equation}
\label{eq:EigProblem}
 \sum_{\bm k'} \mathcal{K}_{\bm k \bm k'} \mathcal{V}_{\bm k' } = 
 \Gamma \mathcal{V}_{\bm k} \, ,
\end{equation}
where $\mathcal{K}_{\bm k\bm k'}$ is the relaxation matrix, defined as 
\begin{equation}
    \label{eq:relmatrix}
    \mathcal{K}_{\bm k\bm k'}/\tau_0=
    - W_{\bm k \bm k^{\prime} }
    + \delta_{\bm k \bm k'}
    \sum_{\bm k ''}W_{\bm k \bm k^{\prime \prime} }
    .
\end{equation} 
Here, we introduce dimensionless time units through a characteristic scattering time, $\tau_0={\hbar}/({\pi n_{\rm imp} U_{\rm imp}^2\rho_0})$, where $\rho_0$ is the electron density of states at energy $\varepsilon_0$. We adopt  $\varepsilon_0=-20$~meV, relative to the valence band top, aligning with natural hole doping in Te~\cite{nanowires}. 

The eigenvector of the relaxation matrix, $\mathcal{V}_{\bm k }$, and its eigenvalue, $\Gamma = \tau_0/\tau$, define a `relaxon' -- a collective relaxation mode with the dimensionless relaxation rate $\Gamma$~\cite{Cepellotti2016, chiral_surfaces}. Relaxons have been mostly used to quantify thermal transport in crystals, where they are collective phonon excitations or wavepackets acting as elementary heat conductors~\cite{Hardy1970, Cepellotti2016,Simoncelli2020}, and recently, to calculate spin relaxation time and diffusion length for free-electron gas with anisotropic Weyl-type SOC~\cite{chiral_surfaces}. In our framework, each relaxon represents a wavepacket of particle-hole excitations above the Fermi see with a well-defined transport lifetime.

The relaxons form a complete orthonormal basis set, evidenced by the completeness relation $\sum_{\bm k } \mathcal{V}_{\bm k}^i \mathcal{V}_{\bm k}^j = \delta_{ij}$, which makes it convenient for finding exact solutions of the Boltzmann equation. Namely, any non-equilibrium electron distribution can be expressed as a linear combination of relaxons~\cite{Hardy1970}:
\begin{equation}\label{eq:specdec}
    \delta f_{\bm k}(t) =   \sum_{i} A_i(t) \, \mathcal{V}_{\bm k}^{\hspace{0.1cm}i} \, ,
\end{equation}
where $i$ labels an individual relaxon, and its amplitude $A_i(t)$ relaxes with the rate $\Gamma_i$ as $A_i(t) = A_i(t=0) e^{-\Gamma_i  \hspace{0.05cm} t/\tau_0}$. Such a spectral decomposition allows us to calculate transport coefficients within the linear regime. For instance, inserting the decomposed $\delta f_{\bm k}$ into the steady-state Boltzmann equation yields the spectral amplitudes of an electrically induced electron population:
\begin{equation}
\label{eq:Ai0}
    A_i(0) = \frac{1}{\Gamma_i}\sum_{\bm k}
    \bm E \cdot \bm{ v}_{\bm k}
    \left(-\frac{\partial f_{\bm k}^{(0)}}{\partial \varepsilon_{\bm k}} \right)
    \mathcal{V}^i_{\bm k} \, ,
\end{equation}
where $\bm{ v}_{\bm k} = \frac{\partial \varepsilon_{\bm k}}{\hbar\partial \bm k}$ is the velocity. For this derivation, we used the completeness relation after multiplying the Boltzmann equation with $\mathcal{V}^j_{\bm k}$ and summing over $\bm k$. Note that we omitted an irrelevant overall factor in Eq.~\eqref{eq:Ai0}, which cancels out the factor $n_{\rm imp} U_{\rm imp}^2$ when complemented by dimensionless time units, making the relaxation time and response coefficients solely dependent on the intrinsic properties of the electronic states, independent of any disorder characteristics. 

Finally, the current-induced spin and orbital angular momentum accumulation can be generally expressed as:
\begin{equation}
    M_a = \chi_{ab} \, j_{b}  \, ,
\end{equation}
where $a,b = x,y,z$ denote the components of the induced magnetization $M_a$ and electron current density $j_b$. The Rashba-Edelstein response tensor is~\cite{EDELSTEIN1990}:
\begin{equation}\label{eq:Mz}
 \chi_{ab}  =  \frac{\sum_{\bm k} \langle \hat{\mu}_a \rangle_{\bm k} \delta f_{\bm k}}
{-e \sum_{\bm k} {v}_{\bm k}^b \delta f_{\bm k}} \, ,
\end{equation}
where $\langle \hat{\mu}_a \rangle_{\bm k}$ is the magnetic dipole moment expectation value for the state $\bm k$, with $\hat{\mu}_a = \left(2 \hat{S}_a + \hat{L}_a\right) \mu_B$. Here, $\hat{S}_a = \sigma_a /2$ is the spin-$1/2$ operator, and $\hat{L}_a = -i \varepsilon_{abc} |p_a \rangle  \langle p_b|$ captures the on-site orbital angular momentum contribution from the atomic $p$ orbitals. Note that the denominator corresponds to generated charge current.  
Based on 
Eqs.~\eqref{eq:specdec}-\eqref{eq:Mz}, we calculate the current-induced accumulation and its lifetime in bulk Te, but our exact Boltzmann framework offers a robust methodology for accurately calculating spin transport coefficients in other materials.

\subsection*{Efficiency of the current-induced spin and orbital angular momentum accumulation in chiral Te crystals}

\begin{figure*}
   \centering
   \includegraphics[width=0.99\linewidth]{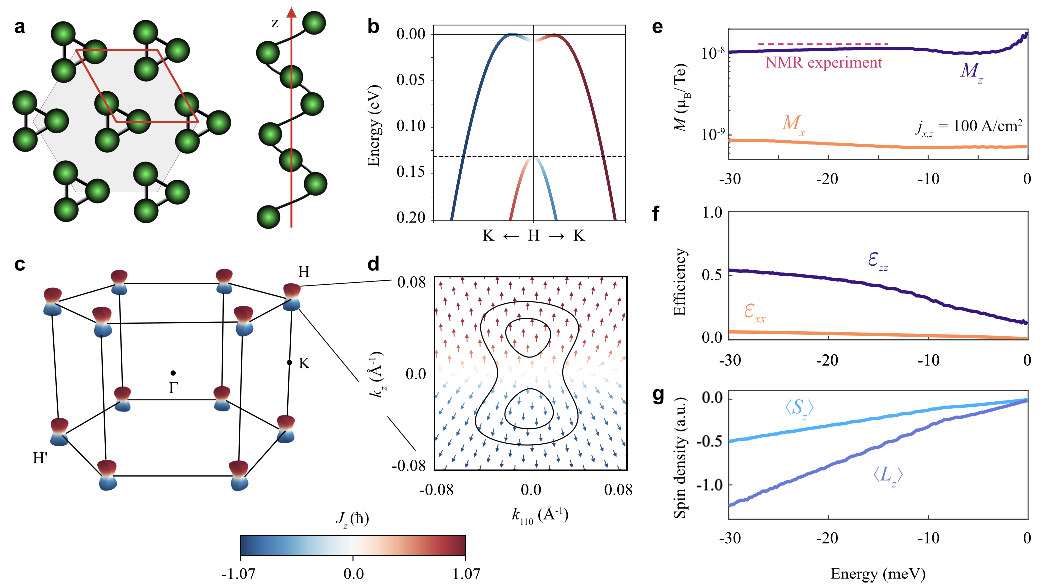}
   \caption{\textbf{$\mid$ Crystal structure, electronic properties and charge-to-spin conversion in tellurium.} {\bf a}  
   Crystal structure of elemental Te. The helicity of an isolated chain along the \textit{z} direction corresponds to right-handed Te. 
   {\bf b} 
   Calculated band structure along the $K H K$ line in the Brillouin zone.
    The color encodes the $z$ component of the total angular momentum. Its $x$ and $y$ components are much smaller in magnitude and are not shown. 
    %
    {\bf c} 
    The three-dimensional Fermi surface at $E = -20$~meV consists of six dumbbell-shaped hole pockets centered in the corners of the hexagonal Brillouin zone. 
    {\bf d}  
    A single Fermi pocket at the $H$ point projected onto a plane parallel to the $z$ axis, where the arrows represent the total angular momentum directions. 
    The constant energy contours at $E= -$ 5~meV (inner-most) and $E= -$20~meV (outer-most) are shown. 
   {\bf e} Current-induced magnetization $M_z$ and $M_x$ per Te atom induced by an electric current along the $z$ and the $x$ axis, respectively, calculated for different values of chemical potential.
   In the realistic doping region, the dashed line indicates the induced magnetization estimated from NMR~\cite{tellurium_main}.
   {\bf f} Charge-to-spin conversion efficiency vs chemical potential. {\bf g} Spin and orbital angular momentum contributions to $M_z$.
   }
\label{fig:EE}
\end{figure*}

Right- and left-handed Te crystals belong to the space symmetry groups $P3_121$ and $P3_221$, respectively. The crystal symmetry enforces current-induced magnetization parallel to charge current, thus allowing only non-zero components $\chi_{zz}$ and $\chi_{xx} = \chi_{yy}$~\cite{analogs}. Because the two enantiomers are connected through inversion, their REE tensors only differ by sign, and we restrict our analysis to the right-handed Te. The details of its structure and calculated electronic properties are shown in Fig.~\ref{fig:EE}{\bf a-d}. The spin- and orbital-resolved electronic states are obtained from density functional theory (DFT) and tight-binding calculations, as described in Methods. 

Figure~\ref{fig:EE}{\bf e} shows the magnetization $M_z$ induced by an electric current along the $z$ axis $j_z = 100$~A~cm$^{-2}$, calculated as a function of the chemical potential. The magnetization value agrees very well with the NMR experimental data, which indicated $M_z = 1.3 \cdot 10^{-8} \mu_{\rm B}$ per Te at $j_z = 82$~A~cm$^{-2}$~\cite{tellurium_main}. We highlight a large improvement compared to the previous theoretical calculations, which underestimated the magnetization by an order of magnitude~\cite{nanowires, arunesh_npj, tellurium_main}. While one could presume a role of the orbital contribution, Fig.~\ref{fig:EE}{\bf g} shows that the orbital contribution alone only doubles the total magnetization, and thus the accuracy primarily comes from incorporating the $k$-state dependent relaxation time, achieved via the exact solution of the Boltzmann equation. 
Notably, electric currents with $j_x = 100$~A~cm$^{-2}$ induce magnetization $M_x$ smaller than $M_z$ by an order of magnitude, showing a strong anisotropy relative to the screw axis. The temperature was set to $T=10$~K for our calculations, but the results are similar at higher temperatures.

Even though the induced magnetization is convenient for comparison with the NMR measurements, for nanodevices that rely on spin transport, the charge-to-spin conversion efficiency is a more important figure of merit. We define the efficiency as:
\begin{equation}
\label{eq:EEepsilon}
\varepsilon_{zz} =   \frac{
\sum_{\bm k} \langle \hat{J}_z  \rangle_{\bm k} \delta f_{\bm k}}
{\sum_{\bm k}  |\delta f_{\bm k}|} ,
\end{equation}
where $\hat{J}_z = \left(\hat{S}_z + \hat{L}_z\right)/J$ is the $z$ component of the total angular momentum polarization operator, normalized with $J = {3}/{2}$, and the electric field is applied along the $z$ axis.
Importantly, $\varepsilon_{zz}$ has a transparent physical interpretation, resembling an intuitive efficiency definition $(N_{\uparrow} - N_{\downarrow})/(N_{\uparrow} + N_{\downarrow})$, where $N_{\sigma}$ is the non-equilibrium deviation of the total number of electrons with spin polarization $\sigma$. Whereas the numerator in Eq.~\eqref{eq:EEepsilon} represents the total angular momentum polarization, its denominator quantifies the magnitude of the current-induced shift in the distribution function. In the quasi-classical picture, ${\sum_{\bm k}  |\delta f_{\bm k}|}$ accounts for the induced imbalance between right-moving and left-moving electron populations in momentum space. The efficiency is normalized from 0 to 1, where the former indicates no spin signal, and the latter describes the ideal situation of maximum accumulation with opposite spins of right- and left-moving electrons.

Figure~\ref{fig:EE}{\bf f} shows the charge-to-spin conversion efficiency of Te calculated as a function of the chemical potential. Tellurium demonstrates exceptional charge-to-spin conversion efficiency, reaching 50\% at realistic hole doping levels that correspond approximately to the chemical potential of $ -20$~meV. This value is much higher than in most known materials with strong SOC~\cite{fert}. We note that lower efficiencies of 20-40\% have been recently reported for Te in the context of possible chirality-induced spin selectivity (CISS) based on spin current calculations in a ballistic transport regime~\cite{droghetti, Yang2024}. However, our approach is more suitable for sample sizes larger than the mean free electron path, as it considers non-ballistic diffusive transport by including disorder.

\begin{figure*}
   \centering
   \includegraphics[width=0.99\linewidth]{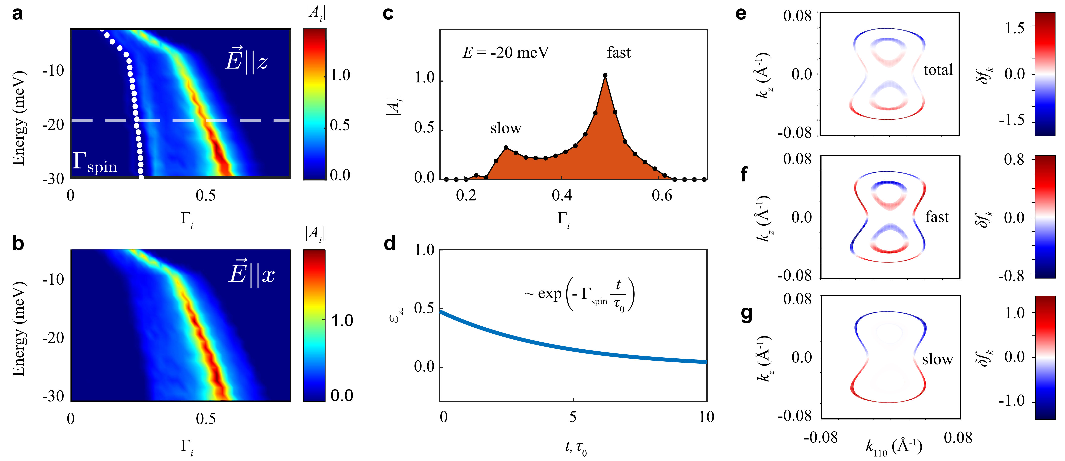}
   \caption{\textbf{$\mid$ Current-induced spin accumulation in Te in the relaxon basis.} 
    {\bf a,b} Spectral decomposition at different values of chemical potential when the electric field $ E$ is applied along the $z$ and the $x$ axis, respectively. The spectral amplitudes are color-coded, and white dots show the calculated spin relaxation rate. For $E\parallel z$, the dashed line corresponds to $E = -20$~meV, where {\bf c} presents its spectrum and {\bf d} its spin accumulation time dependence. {\bf e} The non-equilibrium deviation of the total distribution function $\delta f_{k}$ calculated near the energies -5~meV (inner contour) and -20~meV (outer contour). {\bf f, g} Contributions to $\delta f_{k}$ from relaxons with $\Gamma_i>0.33$ and $\Gamma_i<0.33$, respectively.  
   }
\label{fig:EEtime}
\end{figure*}

\subsection*{Spin accumulation lifetime}
We now examine the time dependence of spin accumulation by analyzing the relaxon spectrum. Figure~\ref{fig:EEtime} presents the spectral decomposition of the current-induced shift in the distribution function. When an electric current is applied along the $z$ axis (see Fig.~\ref{fig:EEtime}{\bf a}), the spectrum has two pronounced peaks corresponding to collective relaxation modes: a fast mode with the relaxation rate $\Gamma \approx 0.5$ and a slow mode with $\Gamma \approx 0.3$ followed by a tiny satellite peak with $\Gamma \approx 0.22$, as shown in Fig.~\ref{fig:EEtime}{\bf c}. 
The contributions to any observable from different collective relaxation modes can be independently calculated using the relaxon spectral decomposition in Eq.~\eqref{eq:specdec}. For example, the average spin density is given by $ S_z(t) = \sum_{\bm k} \langle \hat{S}_z \rangle_{\bm k} \delta f_{\bm k}(t)/V = \sum_{i} \langle \hat{S}_z \rangle_{i} A_i(t)$, where 
$\langle \hat{S}_z \rangle_{i} = \sum_{\bm k} \langle \hat{S}_z \rangle_{\bm k}\mathcal{V}_{\bm k}^{\hspace{0.1cm}i}/V$ is the spin polarization of the $i$-th relaxon.

We find that the main contribution to $M_z$ comes from the relaxons belonging to the slow modes (that is, $\Gamma_i<0.33$). This is further confirmed by the spin relaxation time, calculated by fitting the spin time dependence to a single exponential decay model (see Fig.~\ref{fig:EEtime}{\bf d}). The resulting spin relaxation rate, marked by white dots in Fig.~\ref{fig:EEtime}{\bf a}, closely follows the slow relaxation modes, demonstrating their dominant role in governing spin accumulation. Conversely, when the electric current flows along the $x$ axis (see Fig.~\ref{fig:EEtime}{\bf b}), the slow modes vanish. The low calculated current-induced magnetization $M_x$ suggests that the fast mode corresponds to the mean electron scattering time, {\it i.e.}, a normal relaxation mode. 

To clarify the origin of the slow mode, we visualize $\delta f_{k}$ on the Fermi pockets near the $H$ point. Figure~\ref{fig:EEtime}{\bf e} displays the $-20$~meV (and $-5$~meV) Fermi contours, where the color quantifies $\delta f_{k}$, and Figs.~\ref{fig:EEtime}{\bf f} and {\bf g} isolate the fast and slow modes by calculating the contributions to $\delta f_{k}$ from relaxons with $\Gamma_i>0.33$ and $\Gamma_i<0.33$, respectively. For $E=-20$~meV, the distribution $\delta f_k$ associated with the slow mode, where the spin polarization of induced holes is opposite to that of electrons, is identical to the angular momentum polarization pattern of electron states in Te (see Fig.~\ref{fig:EE}{\bf d}). This alignment of the slow mode distribution $\delta f_{\bm k}$ with the spin polarization texture in $\bm k$-space reduces back-scattering into the inner part of the dumbbell-shaped pocket, thereby slowing this mode's relaxation. Since non-magnetic impurity scattering does not flip spin, the scattering probability $W_{\bm k \bm k'}\sim |\langle \bm k' | \bm k \rangle|^2$ is diminishing when the states $|\bm k \rangle$ and $|\bm k' \rangle$ have opposite spin polarizations. Nonetheless, the suppression is only partial because the spin polarization of the electron states is partial. The total angular momentum of these states never attains its maximum value of $3/2$, as illustrated in Fig.~\ref{fig:EE}{\bf d}, in agreement with the angle-resolved photoemission spectroscopy (ARPES) results~\cite{sakano}. The contour at $E=-5$~meV shows a different scenario, where the slow mode is nearly absent in the spectral decomposition (see Fig.~\ref{fig:EEtime}{\bf g}), resulting in much smaller current-induced accumulation (Fig.~\ref{fig:EE}{\bf g}) and shorter spin lifetime (Fig.3{\bf a}).

\subsection*{Low-energy model and persistent spin helix}

The presence of slow relaxons in chiral tellurium crystals, which lead to efficient spin accumulation with extended lifetime, can be generalized to other systems.
To study this, we examine the relationship between the spin texture of the valence band and the slow relaxation mode using a $ k\cdot p$ model that describes the low-energy behavior of holes in Te~\cite{old1,old2,old_farbshtein}. We show that this slow mode corresponds to the persistent spin helix, similar to those observed in GaAs quantum wells~\cite{koralek}. 

The two upper-lying valence bands are described by the following two-state Hamiltonian:
\begin{equation}\label{eq:kp_model}
    \hat{\mathcal{H}}_k = -A k_z^2 - B \left(k_x^2 + k_y^2\right) +
    \beta k_z \hat{\tau}_z + \Delta \hat{\tau}_x,
\end{equation}
where $\hat{\tau}_z$ and $\hat{\tau}_x$ are Pauli matrices in a pseudospin space.
This model resembles the free electron gas model with strongly anisotropic Weyl-type SOC, subject to an in-plane magnetic field indicated by the Zeeman gap $\Delta$.
It is minimal for describing the opening of the energy gap between the two valence bands, as shown in  Fig.~\ref{fig:EE}{\bf b}, and the Hamiltonian becomes diagonal in the basis of states $|\psi_\pm\rangle$, where $|\psi_+\rangle$ describes the upper band, and $|\psi_-\rangle$ describes the lower band.

While detailed model parameters, band dispersion, and wavefunctions are discussed in Methods, it is crucial to note that the pseudospin polarization, as given by $\langle\psi_+| \hat{\tau_z} |\psi_+\rangle \sim \beta k_z$, largely reflects the spin polarization, making the pseudospin quantum number useful for interpreting the slow relaxation mode in connection with the spin texture of the electron states. 
Next, we will solve Boltzmann transport equations to generalize the concept of slow relaxons.

\begin{figure}
   \centering
   \includegraphics[scale=1.0]{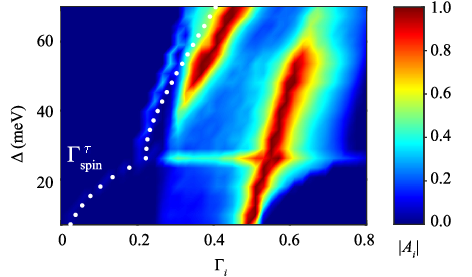}
   \caption{\textbf{$\mid$ Current-induced pseudospin accumulation in Te calculated using the low-energy model.} 
    Spectral decomposition at different values of $\Delta$. The electric field is applied along the $z$ axis. The spectral amplitudes are color-coded, and white dots show calculated pseudospin relaxation rate.  
    The spectrum at each value of $\Delta$ is normalized to the unit at its maximum value.
   The Fermi level is at $E_{\rm F} = -20$~meV.
   }
\label{fig:kp_model}
\end{figure}

The relaxon spectral decomposition of the REE shows a strong dependence on $\Delta$.
Figure~\ref{fig:kp_model} shows the impact of varying the magnitude of $\Delta$ while keeping a constant chemical potential of $E = -20$~meV. 
By analogy with the previous section, we calculate the average pseudospin value, $\sum_{\bm k} \langle \hat{\tau}_z \rangle_{\bm k} \delta f_{\bm k}(t)/V$, and its relaxation time. For realistic values describing Te bands, with $\Delta \approx 60$~meV, the pseudospin relaxation rate and the two relaxation peaks are in agreement with those shown in Fig.~\ref{fig:EEtime}{\bf a}. 
As $\Delta$ decreases, the spectral weight of the slow mode diminishes and nearly disappears at $\Delta \approx 25$~meV. This threshold matches an abrupt change in the Fermi contours topology from a dumbbell shape (see  Fig.~\ref{fig:EE}{\bf d}) to two separate circles. When the neck of the dumbbell shrinks to zero, the fast relaxation mode dominates.

The pseudospin relaxation rate, indicated by white dots in Fig.~\ref{fig:kp_model}, closely follows a third tiny peak in the spectrum. This third peak, a satellite of the slower mode, is visible in the $\Delta$-region between 30 and 60 meV. Although the slow mode disappears through the crossover region at $\Delta \approx 25$, the pseudospin polarization does not completely disappear but gradually reduces as $\Delta$ decreases.
The REE diminishes and ultimately disappears in the small $\Delta$ limit, as predicted for uniaxial spin textures in GaAs~\cite{Duckheim2010,Trushin2007}.
Remarkably, the pseudospin relaxation rate eventually drops to zero as $\Delta$ approaches small values.

These observations indicate that the slow mode depicted in Fig.~\ref{fig:kp_model} tracks the $\Delta$-evolution of the persistent spin helix state.
If $\Delta$ were exactly zero, our model would be equivalent -- up to a rotation of coordinates -- to the system with equal Rashba and Dresselhaus parameters, where the PSH has been observed. There is, however, one important difference: in GaAs, the spins and the shift in momentum space are orthogonal to each other, whereas in the Weyl-type SOC, they are parallel, aligning the magnetization induced in Te with the applied current.
Despite this difference, the argument suggesting emergent SU(2) spin rotational symmetry~\cite{berenevigpst} still applies. It suggests that the PSH state manifests as a spiral spin state, where the spins lie in the plane that is orthogonal to the PSH wavevector. Conversely, a finite $\Delta$ breaks the SU(2) symmetry, introducing non-collinearity in the momentum space spin texture and thereby imposing a finite lifetime of the spin helix. Thus, we conclude that the slow mode corresponds to the PSH, whose finite lifetime is due to the non-zero gap $\Delta$.

\subsection*{Final remarks}

In summary, we reveal the role of the slow relaxation modes in creating highly efficient and long-living current-induced spin accumulation in bulk crystals of chiral Te. Although Te has been known and studied for decades, our results show it from a new perspective; on the one hand, it allows for highly efficient creation of spin density by applied electric current -- a feature of materials with strong SOC, and on the other hand, it promises long-range spin diffusion -- a feature of materials with weak SOC, such as graphene. Such a combination of properties makes it ideal for novel spintronic applications. For example, an all-electrical tellurium-based device could operate in the non-local spin diffusive regime: in an `injector' region, an electric current applied along the $z$ axis induces magnetization along the $z$ axis; in a spin `conductor' region, the spin starts diffusing over hundreds of nanometers; in a `detector' region, the spin signal is measured electrically through spin-to-charge conversion. Our approach and results lay the groundwork for the search for other materials, where slow relaxons can generate and carry spin accumulation over long distances.

\section*{Methods}

\subsection*{First-principles calculations for Te}
We performed DFT calculations for bulk Te using the Quantum Espresso package~\cite{qe1,qe2}. We employed the Perdew, Burke, and Ernzerhof generalized gradient approximation for exchange-correlation functional as well as fully relativistic pseudo-potentials~\cite{pslibrary, pbe, kresse-joubert}. The electron wave functions were expanded in a plane-wave basis with the energy cutoff of 80 Ry. The structure with a hexagonal unit cell containing three atoms and lattice constants $a$=4.52 \AA\  and $c$ = 5.81 \AA\ was adopted after the full optimization with the energy and force convergence criteria set to 10$^{-5}$ Ry and 10$^{-4}$ Ry/bohr, respectively. The BZ was sampled following the Monkhorst-Pack scheme with the $k$-grids of 22$\times$22$\times$16 and a Gaussian smearing of 0.001 Ry. The electronic structure was additionally corrected with the Hubbard parameter ($U_{5p}$ = 3.81 eV) calculated self-consistently via the ACBN0 approach~\cite{acbn0}. The spin-orbit interaction was included self-consistently in the calculation. As a post-processing step, we used the open-source Python code PAOFLOW~\cite{paoflow1,paoflow2} to project DFT wave functions onto the pseudo-atomic orbitals and construct tight-binding Hamiltonians~\cite{agapito0,agapito1, agapito2}. We further interpolated these Hamiltonians to denser $k$-grids of 101$\times$101$\times$101 around the $H$ high-symmetry point of the BZ ($k_z \in k_z^{H}\pm 0.06 \cdot2 \pi/c$ and $k_{x,y} \in k_{x,y}^{H} \pm 0.03\cdot 2 \pi/a$) to accurately calculate the response coefficients.  

\subsection*{Boltzmann transport equations}

The \textit{ab initio} wave functions and energy bands in the pseudoatomic basis are used to construct the relaxation matrix $\mathcal{K}_{\bm k\bm k'}$, defined in Eq.~\eqref{eq:relmatrix}. 
The relaxation matrix incorporates the scattering probability,  $W_{\bm k \bm k^{\prime} }$, which encodes the spin, orbital, and sublattice degrees of freedom within the scattering amplitudes. These amplitudes are given by 
$
    \langle \bm k' | H_{\rm int} | \bm k \rangle =
    \frac{U_{\rm imp}}{V} \sum_a 
    e^{-i\left(\bm k' - \bm k \right)\cdot \bm r_a}
    \langle \bm k' | \bm k \rangle
    $,
where $V$ is the total volume. To obtain the scattering probability using Fermi's golden rule, we calculate the square of the matrix elements as:
$
|\langle \bm k' | H_{\rm int} | \bm k \rangle|^2
 =
    \frac{U_{\rm imp}^2 n_{\rm imp}}{V} 
    |\langle \bm k' | \bm k \rangle|^2.
$ 
We regularize the Dirac delta function using the Lorentzian form:
$
\delta\left(\varepsilon_{\bm k'} - \varepsilon_{\bm k}\right)=
\frac{1}{\pi}\frac{\Delta_E}{ \Delta_E^2 +\left(\varepsilon_{\bm k'} - \varepsilon_{\bm k}\right)^2 }$,
where the width $\Delta_E$ is set to 0.2~meV in our calculations. This regularization is also used to calculate the density of states:
$
\rho_0 = 
\frac{1}{V} \sum_{\bm k} \delta \left(\varepsilon_{0} - \varepsilon_{\bm k}  \right)
$.
To diagonalize the relaxation matrix for each calculation at a given Fermi level, we include electron states in a small window near the Fermi level ($\pm 2$~meV). 
By numerically diagonalizing the relaxation matrix, we obtain a set of eigenvalues, $\Gamma_i$, and eigenstates, $\mathcal{V}_{\bm k}^i$, which form the relaxon basis. Using this basis, we calculate the spectral amplitudes $A_i(0)$, as defined in Eq.~\eqref{eq:specdec}, to determine the variation in the distribution function $\delta f_k$ due to an electric current.

\subsection*{Low-energy Hamiltonian near the $H$ point}

Diagonalizing the Hamiltonian in Eq.~\eqref{eq:kp_model} gives the upper and lower valence band dispersions:
\begin{equation}\label{eq:kp_energy}
    E_{\pm} = -A k_z^2 - B \left(k_x^2 + k_y^2\right) \pm \sqrt{
    \Delta^2 +\beta^2 k_z^2
    }
    ,
\end{equation}
described by the wavefunctions:
\begin{equation}
|\psi_\pm\rangle =\left(\sqrt{1+\chi} |1/2 \rangle \pm \sqrt{1-\chi}|-1/2 \rangle \right)/\sqrt{2} 
\end{equation}
with $\chi(k_z) = \beta k_z/\sqrt{\Delta^2 + \beta^2 k_z^2}$.
The pseudospin polarization of the upper band, given by
$\langle\psi_+| \hat{\tau_z} |\psi_+\rangle = \chi$,
 closely follows the spin polarization shown in Fig.~\ref{fig:EE}{\bf d}.
Note that the pseudospin is often referred to as $| \pm 3/2 \rangle$ states in the literature, convenient for multiband representations described by higher effective spins. This different labeling does not affect our calculations. The resulting upper and lower band wavefunctions, $|\psi_+\rangle$ and $|\psi_-\rangle $, transform as irreducible representations of the symmetry group of the $H$ point: $H_4$ and $H_5$, respectively.
For the transport calculations, we use the following $k\cdot p$ model parameters: $A=32.6$~eV~\r{A}$^2$, $B=36.4$~eV~\r{A}$^2$, $\beta=2.47$~eV~\r{A} and $\Delta=63$~meV~\cite{old_farbshtein, tellurium_main}.
In these calculations, the disorder matrix elements are diagonal in the pseudospin space, although in reality, they are diagonal in the spin space. 
Nevertheless, we expect the results to qualitatively agree with those obtained from the \textit{ab initio} wavefunctions, given that the pseudospin polarization closely follows the spin and orbital angular momentum polarization of Te bands.

\section*{Acknowledgments}
\noindent We are grateful to B.J. van Wees, M. Mostovoy, K. Sundararajan, S. Tirion, S. B. Kilic, and F. Cerasoli for the helpful discussions. J.S. acknowledges the Rosalind Franklin Fellowship from the University of Groningen. E.B. and J.S. acknowledge the grant of the Dutch Research Agenda (NWA) under the contract NWA.1418.22.014 financed by the Dutch Research Council (NWO). The calculations were carried out on the Dutch national e-infrastructure with the support of SURF Cooperative (EINF-5312) and on the H\'{a}br\'{o}k high-performance computing cluster of the University of Groningen.

\section*{Data availability}
\noindent The data associated with the manuscript will be available after the publication via DataverseNL. 

\section*{Code availability}

\noindent The MATLAB code used in this study to obtain exact solutions of the Boltzmann equation is available at GitHub (\url{https://github.com/EBarts/Boltzmann_Solver}).


\bibliographystyle{apsrev4-2} %
\bibliographystyle{naturemag}
\bibliography{TePST_lib} 

\end{document}